\documentclass[a4paper]{jpconf}
\usepackage{graphicx}
\begin{document}
\title{Neutron rich matter, neutron stars, and their crusts}

\author{C. J. Horowitz}

\address{Nuclear Theory Center and Department of Physics, Indiana University, Bloomington, IN 47405, USA}

\ead{horowit@indiana.edu}

\begin{abstract}
Neutron rich matter is at the heart of many fundamental questions in Nuclear
Physics and Astrophysics.  What are the high density phases of QCD? Where did the chemical elements come from? What is the structure of many compact and energetic objects in the heavens, and what determines their electromagnetic, neutrino, and gravitational-wave radiations?  Moreover, neutron rich matter is being studied with an extraordinary variety of new tools such as  Facility for Rare Isotope Beams  (FRIB) and the Laser Interferometer Gravitational Wave Observatory (LIGO).  We describe the Lead Radius Experiment (PREX) that is using parity violation to measure the neutron radius in 208Pb. This has important implications for neutron stars and their crusts. Using large scale molecular dynamics, we model the formation of solids in both white dwarfs and neutron stars. We find neutron star crust to be the strongest material known, some 10 billion times stronger than steel.  It can support mountains on rotating neutron stars large enough to generate detectable gravitational waves.  Finally, we describe a new equation of state for supernova and neutron star merger simulations based on the Virial expansion at low densities, and large scale relativistic mean field calculations.
\end{abstract}

\section{Introduction}

Neutron rich matter is at the heart of many fundamental
questions in nuclear physics and astrophysics.
\begin{itemize}
\item What are the high density phases of QCD?
\item Where did the chemical elements come from?
\item What is the structure of many compact and energetic
objects in the heavens, and what determines their
electromagnetic, neutrino, and gravitational-wave
radiations?
\end{itemize}
Furthermore, neutron rich matter is being studied with an extraordinary
variety of new tools such as the Facility for Rare Isotope
Beams (FRIB), a heavy ion accelerator to be built at
Michigan State University, and the Laser Interferometer
Gravitational Wave Observatory (LIGO).  Indeed there are many, qualitatively different, probes of neutron rich matter including precision laboratory experiments on stable nuclei and experiments with neutron rich radioactive beams.  While astrophysical observations probe neutron rich matter with electromagnetic radiation, neutrinos, and gravitational waves.  Finally, there are new theoretical probes of neutron rich matter that have been enabled by tremendous advances in computational science.  In this paper we give brief examples of how neutron rich matter is being studied with some of these different probes.

\section{Lead Radius Experiment (PREX)}

We start with the Lead Radius Experiment (PREX) \cite{prex}.  This is a precision experiment to measure the neutron radius in $^{208}$Pb with parity violating electron scattering \cite{bigpaper}.  This has many implications for nuclear structure, astrophysics, atomic parity violation, and low energy tests of the standard model.   

Parity violation provides a model independent probe of neutrons, because the $Z^0$ boson couples to the weak charge, and the weak charge of a proton $Q_W^p$,
\begin{equation}
Q_W^p=1-4\sin^2\Theta_W \approx 0.05,
\label{qwp}
\end{equation}
is much smaller than the weak charge of a neutron $Q_W^n$,
\begin{equation}
Q_W^n=-1.
\end{equation}
In Eq. \ref{qwp}, $\Theta_W$ is the weak mixing angle.  Therefore, the weak neutral current interaction, at low momentum transfers, couples almost completely to neutrons.

PREX will measure the parity violating asymmetry $A_{pv}$.  This is the fractional cross section difference for scattering positive (+), or negative (-), helicity electrons,
\begin{equation}
A_{pv}=\frac{\frac{d\sigma}{d\Omega}|_+-\frac{d\sigma}{d\Omega}|_-}{\frac{d\sigma}{d\Omega}|_++\frac{d\sigma}{d\Omega}|_-}\, .
\end{equation}
In Born approximation, $A_{pv}$ arrises from the interference of a weak amplitude of order $G_F q^2$, with $G_F$ the Fermi constant and $q$ the momentum transfer, and an electromagnetic amplitude of order the fine structure constant $\alpha$ \cite{sick},
\begin{equation}
A_{pv}\approx \frac{G_F q^2\, F_W(q^2)}{2 \pi\alpha\sqrt{2} \, F_{ch}(q^2)}\, .
\label{apvborn}
\end{equation}
Here the weak form factor $F_W(q^2)$ is the Fourier  transform of the weak charge density $\rho_W(r)$, that is essentially the neutron density,
\begin{equation}
F_W(q^2)=\int d^3r \frac{\sin(qr)}{qr} \rho_W(r)\, .
\end{equation}
Likewise, the electromagnetic form factor $F_{ch}(q^2)$ is the Fourier transform of the (electromagnetic) charge density $\rho_{ch}(r)$.  This is known from elastic electron scattering \cite{pbden}. Therefore, in principle, measuring $A_{pv}$ as a function of $q$ allows one to map out the neutron density $\rho_n(r)$.  Note that, for a heavy nucleus, there are important corrections to Eq. \ref{apvborn} from Coulomb distortions.  However, these have been calculated exactly by solving the Dirac equation for an electron moving in both a Coulomb potnetial of order 10 MeV and a weak axial vector potential of order electron volts \cite{coulombdistortions}.  Therefore, even with Coulomb distortions, one can accurately determine neutron densities.   Note that this purely electroweak reaction is free from most strong interaction uncertainties.

The PREX experiment in Hall A at Jefferson Laboratory measures $A_{pv}$ for 1.05 GeV electrons elastically scattered from $^{208}$Pb at laboratory angles near five degrees.  The experiment uses a 0.5 mm thick Pb foil target that is enriched in $^{208}$Pb and sandwiched between two thin diamond foils. These diamond foils help remove the beam heating and keep the target from melting.  Scattered electrons are deflected by septum magnets into the two high resolution spectrometers that separate elastically from inelastically scattered electrons.  The goal of this challenging experiment is to measure $A_{pv}$, that is of order 0.5 ppm, to 3\%.   This allows one to determine the neutron root mean square radius to 1\% ($\pm 0.5$ fm).

The experiment ran during April to June 2010.  Krishna Kumar, one of the PREX spokespersons, provided the following preliminary results \cite{kk}.  The experiment was successfully commissioned in March/ April.  High quality data were accumulated in transverse spin mode so as to make the systematic error from a potnetial 2 photon exchange amplitude negligible.  However, radiation issues downstream from the main apparatus resulted in a significant reduction in experiment ``up time''.   Sufficient statistics were accumulated to provide a significant first constraint on the neutron radius.  The collaboration purposes to design appropriate engineering modifications to the beamline to mitigate the radiation problem and request additional beam time to achieve the original goal of a 1\% constraint on the neutron radius.

\section{Neutron star radii}

If one can accurately measure the neutron radius with PREX, then there will be important implications for nuclear structure, atomic parity violation \cite{bigpaper}, and astrophysics.  Let us discuss some of the implications for astrophysics.  Heavy nuclei are expected to have a neutron rich skin.  The pressure of neutron matter in this skin forces neutrons out against surface tension.  Therefore the larger the pressure of neutron matter, the larger the neutron radius in $^{208}$Pb \cite{abrown}.  Measuring the neutron radius of $^{208}$Pb allows us to determine the equation of state of neutron matter (pressure as a function of density) for slightly subnuclear densities near 0.1 fm$^{-3}$.  This density represents an average of the surface and interior densities of $^{208}$Pb.

There is a very interesting relationship between the neutron radius of $^{208}$Pb, of order 6 fm, and the radius of a neutron star, of order 10 km \cite{pbversusNS}.  This involves a breathtaking extrapolation of 18 orders of magnitude in size, or 55 orders of magnitude in mass.  Nevertheless, both in the laboratory and in astrophysics, it is the same neutrons, the same strong interactions, the same neutron rich matter, and the same equation of state.  A measurement in one domain can have important implications in the other domain.  
     
The radius of a neutron star depends on the pressure of neutron matter at normal nuclear density and above, because the central density of a neutron star can be a few or more times that of normal nuclear density.   It is important to have both low density information on the equation of state from PREX, and high density information from measurements of neutron star radii.  This can constrain any possible density dependence of the equation of state from an interesting phase transition to a possible high density exotic phase such as quark matter, strange matter, or a color superconductor.  For example, if the $^{208}$Pb radius is relatively large, this shows the EOS is stiff at low density (has a high pressure).  If at the same time, neutron stars have relatively small radii, than the high density EOS is soft with a low pressure.  This softening of the EOS with density could strongly suggest a phase transition to a soft high density exotic phase.

The radius of a neutron star $R$ can be deduced from X-ray measurements of luminosity $L$ and surface temperature $T$,
\begin{equation}
L=4\pi R^2 \sigma_{SB}T^4,
\label{NSR}
\end{equation}       
with $\sigma_{SB}$ the Stefan Boltzmann constant.  This allows one to deduce the surface area of a neutron star.  However, there are important complications.  First one needs an accurate distance to the neutron star, for example from optical parallax measurements.   Second, Eq. \ref{NSR} assumes a black-body and there are important non-blackbody corrections that must be determined from models of neutron star atmospheres.  Finally, gravity is so strong that space near a neutron star is strongly curved.  If one looks at the front face of a neutron star, one will also see about 30\% of the far face, because of the curvature of space.  Therefore the surface area that one observes in Eq. \ref{NSR} depends on the curvature of space and the mass of the star.  Thus what started out as a measurement of just the radius is, in fact, a measurement of a combination of mass and radius. 

Recently Steiner, Lattimer, and Brown have deduced masses and radii \cite{SLB} from combined observations of seven neutron stars in three classes (1) X-ray bursts, (2) neutron stars in globular clusters, and (3) the nearby isolated star RXJ 1856.  They conclude that observations favor a stiff high density equation of state that can support neutron stars with a maximum mass near 2 $M_\odot$ and that the equation of state is soft at low densities so that a 1.4 $M_\odot$ neutron star has a radius near 12 km.  They go on to predict that their equation of state, fit to observations of neutron stars, implies the neutron minus proton root mean square radii in $^{208}$Pb will be $R_n-R_p=0.15\pm 0.02$ fm.

The Steiner et al. paper \cite{SLB} is potentially controversial because their results depend on, among other things, the model assumed for X-ray bursts.  Ozel et al. use a different model for X-ray bursts, with more optimistic assumptions about the errors, and get very small neutron star radii near 10 km or below \cite{ozel}.   Clearly these determinations of neutron star masses and radii, if correct, have very important implications for the properties of cold dense QCD.

The neutron radius of $^{208}$Pb has implications for neutron star structure, in addition to the star's radius.   Neutron stars have solid crusts over liquid cores, while a heavy nucleus is expected to have a neutron rich skin.  Both the solid crust of the star, and the skin of the nucleus, are made of neutron rich matter at similar, slightly subnuclear, densities.   The common unknown is the equation of state of neutron matter.   A thick neutron skin in $^{208}$Pb, means a high pressure where the energy rises rapidly with density.  This quickly favors the transition to a uniform liquid phase.  Therefore, we find a strong correlation between the neutron skin thickness, measured by PREX, and the transition density in neutron stars from solid crust to liquid interior \cite{crustvsskin}.     

\section{How stars freeze}

Perhaps the presence of a solid crust deserves comment.  We expect solids to form in compact stars, both white dwarfs and neutron stars.  This may sound surprising since stars are often composed of not liquids or gasses, but plasmas.  Nevertheless, the plasmas can be so dense that the ions actually freeze.  Recently, there are significant new observations of how white dwarfs freeze \cite{winget} and of how the solid crust of accreting neutron stars cools \cite{crustcooling,crustcooling1,crustcooling2}.   We hope to extract important new information, on the structure of solids in stars, by comparing these observations to our large scale molecular dynamics simulations of freezing in both white dwarfs \cite{wdprl} and neutron stars \cite{phasesep}.

\subsection{White dwarf crystallization}
We start with white dwarf crystallization.  Winget et al. deduce the luminosity function for white dwarfs in the globular star cluster NGC 6397 \cite{winget}, see also \cite{garcia-berro}.  This is simply the number of white dwarfs in different luminosity bins.  As white dwarfs cool, their luminosities decrease.  The probability to find a white dwarf with a given luminosity depends on the cooling rate.  One is more likely to find stars at luminosities where they cool slowly, and hence spend a long time.  Therefore, the luminosity function is expected to be proportional to one over the cooling rate.   Eventually as a star cools, it freezes.  When this happens, the latent heat will slow the cooling rate until finally all of the latent heat is radiated away.  This is exactly like ice cubes slowing the rate of warming of your drink.  Therefore there should be a peak in the luminosity function when the stars freeze.  Indeed such a peak is observed by Winget et al. \cite{winget}.  Furthermore, the luminosity where this peak occurs, depends on the interior temperature.  This allows one to deduce the melting temperature of white dwarf cores.          

We expect the cores of the white dwarfs in NGC 6397 to be a mixture of carbon and oxygen.  From the observed melting temperature one can infer the ratio of carbon to oxygen, if one knows the phase diagram for carbon / oxygen mixtures.  Indeed the ratio of carbon to oxygen in white dwarf cores is important for stellar evolution and nucleosynthesis.  It depends on the rate for the, very interesting but incompletely known, nuclear reaction $^{12}$C($\alpha,\gamma$)$^{16}$O.  Note that there have been previous efforts to determine the composition of white dwarf cores using astrosysmology, where the frequencies of different oscillation modes are observed \cite{WDsysmology}.  However, the interpretation of these frequencies may be ambiguous.  

\begin{figure}[h]
\center\includegraphics[width=5.2in]{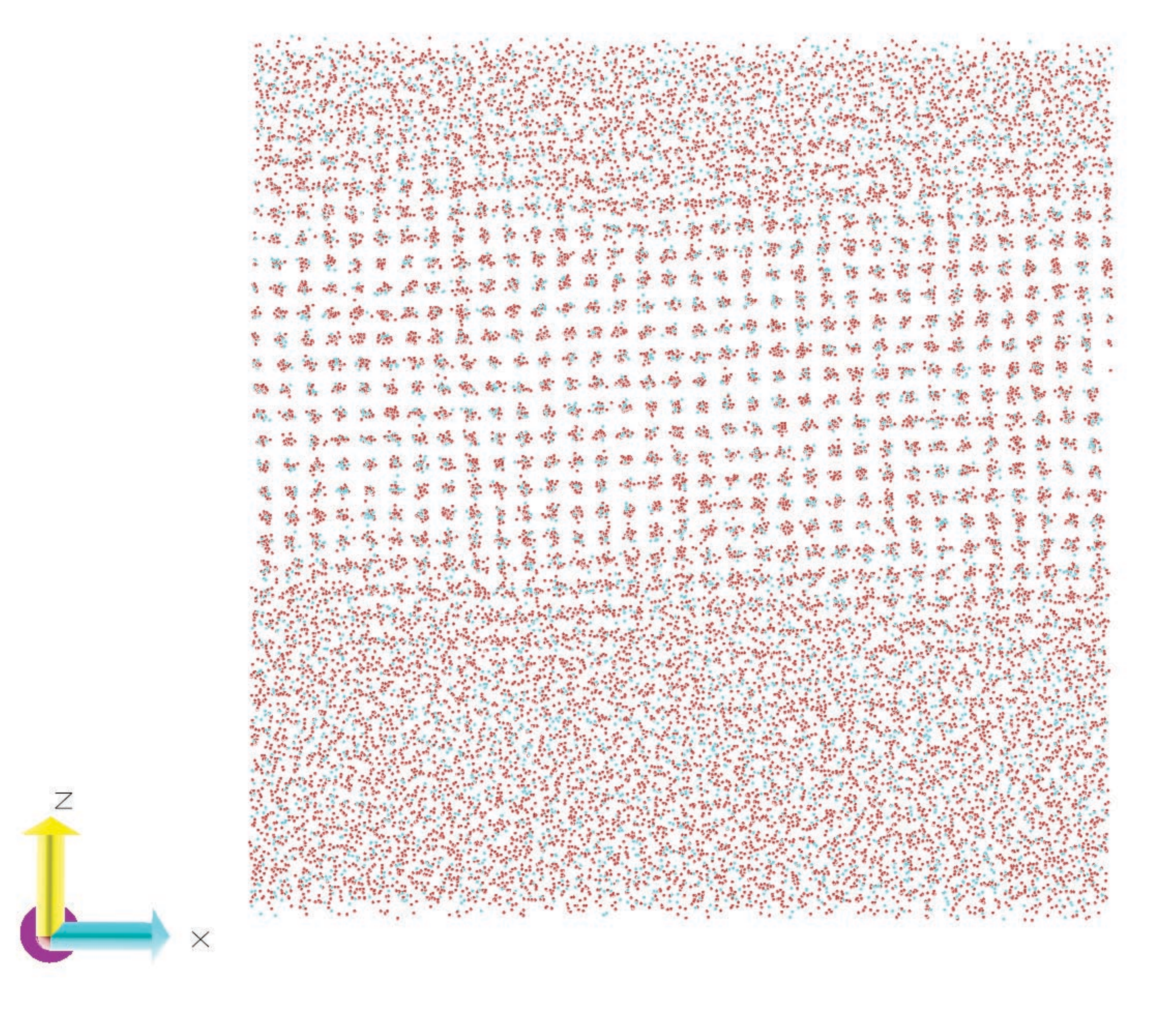}
\caption{\label{Fig1}Molecular dynamics simulation of a 27648 ion carbon and oxygen mixture showing a solid phase (center) in equilibrium with a liquid phase (top and bottom) \cite{wdprl}.}
\end{figure}

To determine the carbon/ oxygen phase diagram, we perform large scale molecular dynamics simulations of the freezing of carbon and oxygen plasma mixtures \cite{wdprl}.  The temperature of a simulation volume containing a 27648 ion carbon / oxygen mixture is carefully adjusted so that half of the system remains solid while the other half is liquid, see Fig. \ref{Fig1}.  Carbon and oxygen ions are free to diffuse between the liquid and solid phases and, in general the solid becomes enriched in oxygen.  We measure the composition of each phase at the end of the simulation.   Because diffusion is slow, these simulations took about six months of computer time on special purpose MDGRAPE-2 boards.  Our phase diagram agrees well with the recent model of Mendin and Cummings \cite{zack}, but gives lower melting temperatures than earlier work by Segratain et al. \cite{segretan}.

Winget et al.'s observations of the white dwarf luminosity function of NGC 6397, together with our carbon/ oxygen phase diagram imply that the oxygen concentration (by number) $x_O$ in the core of these white dwarfs is $x_O<0.57$ \cite{wdprl}.  The oxygen concentration in the core of a white dwarf depends on the rate of the nuclear reaction $^{12}$C($\alpha,\gamma$)$^{16}$O and on the treatment of diffusion in stellar evolution models.  For Salaris et al.'s models of 0.54 $M_\odot$ white dwarfs, the constraint $x_O<0.57$ corresponds to a constraint on the effective astrophysical $S$ factor at 300 keV, describing the  $^{12}$C($\alpha,\gamma$) reaction, of $S_{300}\leq$ 170 keV barns.  This is consistent with the recent experimental determination of Buchmann and Barnes of $S_{300}\approx 145$ keV barns \cite{buckmann}.  Observations of crystallization in many more white dwarf systems is feasible and this may provide important benchmarks for both the $^{12}$C($\alpha,\gamma$) reaction, and our understanding of how nuclear reactions in stars make the carbon in our bones and the oxygen that we breathe.

\subsection{Crystallization in accreting neutron stars}
We next consider freezing in accreting neutron stars.  Material falling on a neutron star can undergo rapid proton capture (or rp process) nucleosynthesis to produce a range of nuclei with mass numbers $A$ that could be as high as $A\approx 100$ \cite{rpash, rpash2}.  As this rp process ash is buried by further accretion, the rising electron fermi energy induces electron capture to produce a range of neutron rich nuclei from O to approximately Se \cite{rpash3}.  This material freezes, when the density reaches near $10^{10}$ g/cm$^3$.  We have performed large scale MD simulations of how this complex rp process ash freezes \cite{phasesep}.  We find that chemical separation takes place and the liquid ocean is greatly enriched in low atomic number $Z$ elements, while the newly formed solid crust is enriched in high $Z$ elements.  Furthermore, we find that a regular crystal lattice forms even though large numbers of impurities are present.  This regular crystal should have a high thermal conductivity.  We do not find an amorphous solid that would have a low thermal conductivity.

Recent X-ray observations of neutron stars, find that the crust cools quickly, when heating from extended periods of accretion stops \cite{crustcooling,crustcooling1,crustcooling2}.  This is consistent with our MD simulations, and strongly favors a crystalline crust over an amorphous solid that would cool more slowly \cite{crustcooling3}.   See also the paper by Ed. Brown in these proceedings \cite{crustcooling4}.  

\section{Gravitational Waves}
We turn now to gravitational waves (GW) from crust ``mountains'' on rapidly rotating neutron stars.  Albert Einstein, almost 100 years ago, predicted the oscillation of space and time known as gravitational waves.  Within a few years, with the operation of Advanced LIGO and other sensitive interferometers, we anticipate their historic detection.  This will be a big deal, after decades of effort, and open a new window on the universe and on neutron rich matter.  

Strong gravitational wave sources often involve large accelerations of large amounts of neutron rich matter.  Indeed the requirements for a strong source of continuous GW, at LIGO frequencies, places extraordinary demands on neutron rich matter.   Generating GW sounds easy.  Place a mass on a stick and shake vigorously.  However to have a detectable source, one may need not only a large mass, but also a very strong stick.   The stick is needed to help produce large accelerations.  Since others have discussed large masses, let us focus here on the strong stick.

Recently we performed large scale MD simulations of the strength of neutron star crust \cite{crustbreaking,chugunov}.   A strong crust can support large deformations or ``mountains'' on neutron stars, see also \cite{lowmassNS}.  These bumps, on rapidly rotating stars, can generate strong gravitational waves.  Indeed Bildstein and others \cite{bildstein} have speculated that some stars in binary systems may spin so fast that they radiate angular momentum in GW at the same rate that angular momentum is gained from accretion.  This would explain why the fastest observed neutron stars are only spinning at about half of the breakup rate. There are several very active ongoing and near future searches for continuous gravitational waves at LIGO, VIRGO and other detectors, see for example \cite{abbot}.  Often one searches at twice the frequency of known radio signals from pulsars because of the quadrupole nature of GW.  Although the signals may be weak, one can gain sensitivity by integrating coherently for long observation times, although this can require very large computational resources.  No signal has yet been detected.  However, sensitive upper limits have been set.  These limits constrain the shape of neutron stars.  In some cases the star's elipticity $\epsilon$, which is that fractional difference in moments of inertia $\epsilon=(I_1-I_2)/I_3$ is observed to be less than a part per million or even smaller.  Here $I_1$, $I_2$, and $I_3$ are the principle moments of inertia.               

\begin{figure}[h]
\center\includegraphics[width=5.6in]{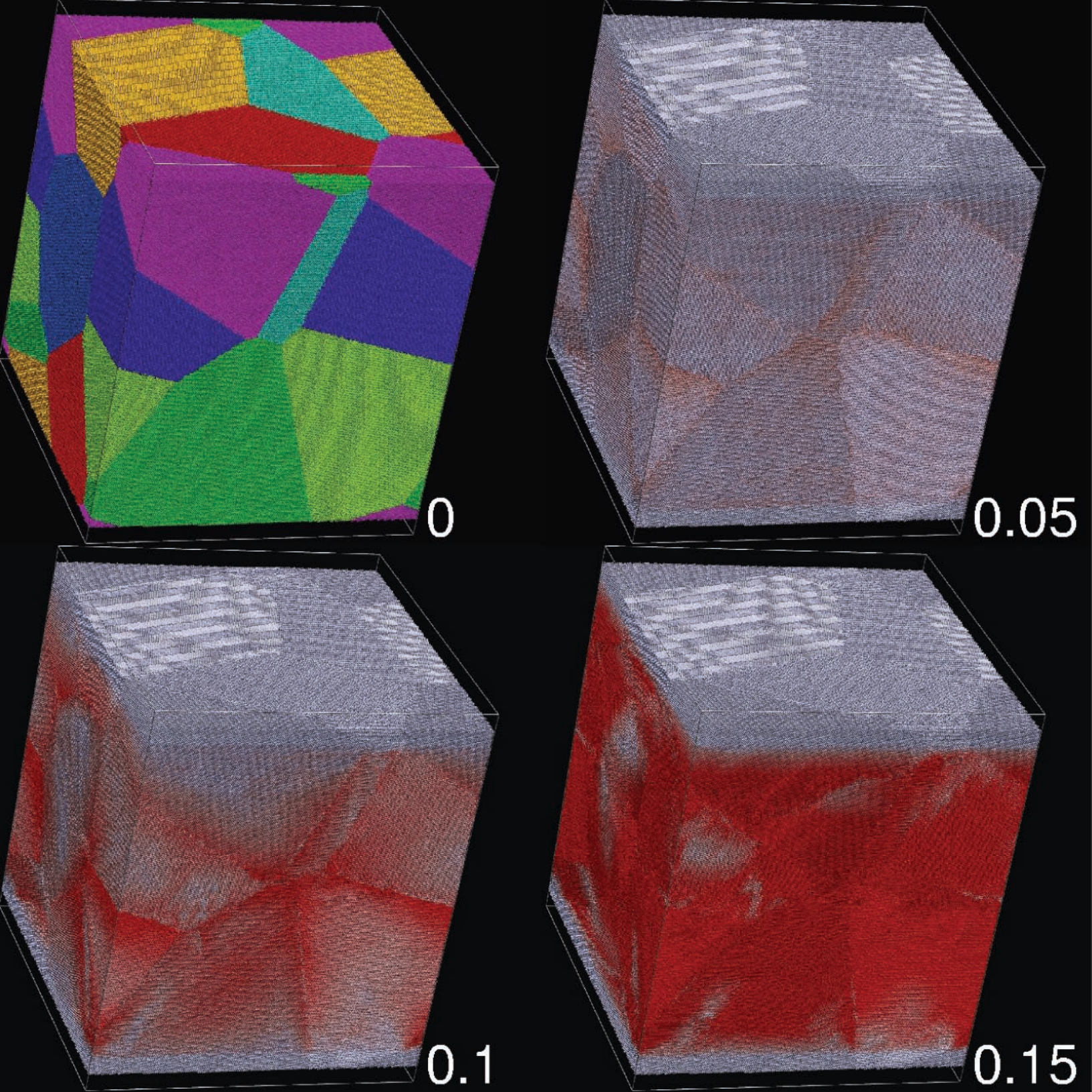}
\caption{\label{Fig2}Molecular dynamics simulation of a 12 million ion polycrystaline sample of neutron star crust.  The four panels are labeled by the strain, which is the fractional deformation.   The colors in the upper left panel, at zero strain, show the original eight microcrystals of different orientations.  The red color in the other three panels indicates distortion of the body centered cubic lattice \cite{crustbreaking}.}
\end{figure}

How large can a neutron star mountain be before it collapses under the extreme gravity?  This depends on the strength of the crust.  We performed large scale MD simulations of crust breaking, where a sample was strained by moving top and bottom layers of frozen ions in opposite directions \cite{crustbreaking}.  These simulations involve up to 12 million ions and explore the effects of defects, impurities, and grain boundaries on the breaking stress.  For example, in Fig. \ref{Fig2} we show a polycrystalline sample involving 12 million ions.  In the upper right panel the initial system is shown, with the different colors indicating the eight original microcrystals that make up the sample.  The other panels are labeled with the strain, i.e. fractional deformation, of the system.  The red color indicates distortion of the body centered cubic crystal lattice.  The system starts to break along grain boundaries.  However the large pressure holds the microcrystals together and the system does not fail until large regions are deformed.

We find that neutron star crust is very strong because the high pressure prevents the formation of voids or fractures and because the long range coulomb interactions insure many redundant ``bounds'' between planes of ions.  Neutron star crust is the strongest material known, according to our simulations.  The breaking stress is 10 billion times larger than that for steel.  This is very promising for GW searches because it shows that large mountains are possible, and these could produce detectable signals.

\section{Neutrinos, supernovae, and neutron rich matter}
Neutrinos provide yet another, qualitatively different, probe of neutron rich matter.  Core collapse supernovae (SN) are gigantic stellar explosions that convert as much as $0.2 M_\odot$ of mass into $10^{58}$ neutrinos.  There are many exciting new underground experiments to search for dark matter, double beta decay, solar neutrinos, proton decay, oscillation of accelerator neutrinos, etc.   Some of these experiments can be very sensitive to neutrinos from the next galactic supernova (SN).  For example, the coming ton-scale dark matter experiments can be sensitive to SN neutrinos via neutrino-nucleus elastic scattering \cite{mckinsey}.  Here, the spectrum of recoiling nuclei provides information of the spectrum of mu and tau neutrinos from a SN that is not available from the existing Super-Kamiokande detector.  Super-K, a large water detector, is primarily sensitive to SN $\bar\nu_e$ via $\bar\nu_e+p\rightarrow n + e^+$.  Note that most of the mass of Super-K is oxygen and this is less sensitive to SN neutrinos than hydrogen.  In contrast, a neutrino-nucleus elastic detector involves very large coherent cross sections, is sensitive to all six flavors of SN neutrinos ($\nu_e,\bar\nu_e,\nu_\mu,\bar\nu_\mu,\nu_\tau$, and $\bar\nu_\tau$) and all of the mass, in the fiducial volume, is active.  This gives very large yields of tens to hundreds of events per ton, instead of per kiloton, for a SN at 10 kiloparsecs \cite{mckinsey}.            

Neutrinos are emitted from a low density, $\approx 10^{11}$ g/cm$^3$, neutron rich neutrino-sphere region.  The pressure, composition, and long wave-length neutrino response \cite{nuresponse} of this region can be calculated from a model independent Virial expansion \cite{virial,virialnuc} that is based on nucleon-nucleon, nucleon-alpha, and alpha-alpha elastic scattering phase shifts.  The abundance of light nuclei near the neutrino-sphere can be important for the radiated $\nu_e$ spectrum \cite{mass3}.  The chance to combine electromagnetic, gravitational wave, and neutrino signals from the next galactic supernova should lead to tremendous results on the dynamics of supernova explosions, neutrinos and neutrino oscillations, the formation of neutron stars, and on nucleosynthesis. 

\section{Equation of state for supernova simulations}
We end with a new equation of state (EOS), pressure as a function of density, temperature, and proton fraction, for neutron rich matter.  This can be used in simulations of SN and neutron star mergers.  At low densities we use a Virial expansion with nucleons, alphas, and thousands of species of heavy nuclei \cite{eos2}.  At high densities we use relativistic mean field calculations in a spherical Wigner-Seitz approximation \cite{eos1}.  Generating the EOS table for a large range of temperatures $T=0$ to 80 MeV, densities $10^{-8}$ to 1.6 fm$^{-3}$, and proton fractions took over 100,000 CPU hours.  This EOS is exact at low density and contains detailed composition information for calculating neutrino interactions.

\section{Conclusions: neutron rich matter}
Neutron rich matter is at the heart of many fundamental questions in Nuclear
Physics and Astrophysics. What are the high density phases of QCD? Where did the chemical elements come from? What is the structure of many compact and energetic objects in the heavens, and what determines their electromagnetic, neutrino, and gravitational-wave radiations? Moreover, neutron rich matter is being studied with an extraordinary variety of new tools such as Facility for Rare Isotope Beams (FRIB) and the Laser Interferometer Gravitational Wave Observatory (LIGO). 

We describe the Lead Radius Experiment (PREX) that is using parity violation to measure the neutron radius in 208Pb. This has important implications for neutron stars and their crusts. Using large scale molecular dynamics, we model neutron rich matter and simulate how material in white dwarfs and neutron stars freeze.   Also, using large scale MD simulations, we find neutron star crust to be the strongest material known, some 10 billion times stronger than steel.  It can support mountains on rotating neutron stars large enough to generate detectable gravitational waves.  Finally, neutrinos from core collapse supernovae provide another, qualitatively different probe of neutron rich matter.  We describe a new equation of state for supernova and neutron star merger simulations based on the Virial expansion at low densities, and on large scale relativistic mean field calculations at high densities.

\ack
This work was done in collaboration with many people including D. K. Berry, E. F. Brown, K. Kadau, J. Piekarewicz, and graduate students Liliana Caballero, Helber Dusan, Joe Hughto, Andre Schneider and Gang Shen.  This work was supported in part by DOE grant DE-FG02-87ER40365, by Shared University Research grants from IBM, Inc. to Indiana University, and by the National Science Foundation, TeraGrid grant TG-AST100014.




\end{document}